\documentclass{iaus}
\usepackage{graphicx}
\def\ang{\AA}
\newcommand{\gapprox}{\lower.4ex\hbox{$\;\buildrel >\over{\scriptstyle\sim}\;$}}
\newcommand{\lapprox}{\lower.4ex\hbox{$\;\buildrel <\over{\scriptstyle\sim}\;$}}
\newcommand{\arcsec}{\hbox{$^{\prime\prime}$}}

\title[Fundamental Physical Processes] %% give here short title %%
{Fundamental Physical Processes in Coronae: Waves, Turbulence, Reconnection,
and Particle Acceleration}

\author[Aschwanden]   
{Markus J. Aschwanden$^1$}
\affiliation{$^1$Laboratory for Solar and Astrophysics, ATC, Lockheed Martin, 
Bldg. 252, Org. ADBS, Palo Alto, CA 94304, USA
\break email: aschwanden@lmsal.com}

\pubyear{2008}
\volume{247}  %% insert here IAU Symposium No.
\pagerange{1-11}
\date{1 October 2008 and in revised form ??}
\jname{Waves \& Oscillations in the Solar Atmosphere: Heating and Magneto-Seismology}
\editors{R. Erdelyi, ed.}
\begin{document}

\maketitle

\begin{abstract}
Our understanding of fundamental processes in the solar corona
has been greatly progressed based on the space observations of SMM,
Yohkoh, Compton GRO, SOHO, TRACE, RHESSI, and STEREO. We observe
now acoustic waves, MHD oscillations, turbulence-related line
broadening, magnetic configurations related to reconnection processes,
and radiation from high-energy particles on a routine basis.
We review a number of key observations in EUV, soft X-rays, and
hard X-rays that innovated our physical understanding of the solar
corona, in terms of hydrodynamics, MHD, plasma heating, and particle
acceleration processes.
%% add here a maximum of 10 keywords from the file <Keywords.txt>
\end{abstract}

\firstsection 

\section{Introduction}

Our grasp to understand the fundamental physical processes in the solar
corona in more depth mostly benefits from the new high-resolution imaging 
capabilities that
became available in extreme ultraviolet (EUV), soft X-rays, hard X-rays, and 
gamma rays. These wavelengths are particularly revealing in the following 
aspects: The corona is filled with high-temperature plasma at temperatures 
of $T\approx 1-2$ MK, which radiates free-free emission with its emission 
measure peak in EUV wavelengths.
Hot active regions and postflare loops have plasma temperatures of
$T\approx 2-40$ MK with their free-free emission peak in soft X-ray
wavelengths. The nonthermal emission of flare-accelerated high-energy
particles radiate bremsstrahlung in hard X-ray wavelengths and have nuclear 
de-excitation lines in gamma rays. Because of the absorbing properties of
the Earth atmosphere, the Sun can be observed in these wavelengths only 
from space.

We discovered waves in the solar corona from oscillatory plasma motions
and from propagating density disturbances. Turbulence 
is somewhat harder to detect, but we mostly infer if from line broadening in 
EUV or from scintillation experiments in radio wavelengths. Magnetic 
reconnection is most conspicuously inferred from the post-reconnection 
topology of post-flare loop systems, but we detect also plasma inflows
and outflows in reconnection regions. Particle acceleration is still the
biggest black box because it occurs on microscopic scales, but we can localize
the acceleration regions from direct bremsstrahlung in the acceleration region
itself, or from reconstructed particle trajectories using time-of-flight
measurements. Here we review the fundamental physics of these four processes,
applied to solar and stellar coronae, and touch on some new challenging 
observations.

\section{Waves in Coronae}

Waves and acoustics have been established as fundamental fields in physics
since the days of Pythagoras, Isaac Newton, and Christian Huygens. Wave
phenomena are governed by an oscillatory dynamics between two competing
forces, such as the inertial motion and the restoring force of a mechanical
elastic medium, or the electric and magnetic field in an electromagnetic wave. 
In the solar corona we have magnetic field lines that are filled with plasma
like fluxtubes, which are governed by magnetohydrodynamics (MHD) and can
exhibit propagating and standing waves. These MHD wave modes can be derived
from the so-called ideal MHD equations, by inserting the current density
${\bf j}=(1/4\pi)\nabla \times {\bf B}$, the electric field ${\bf E}=
(-1/c) {\bf v} \times {\bf B}$ from Ohm's law, the definition of the
sound speed $c_S = \gamma p / \rho$, and assuming an adiabatic process
with  polytropic index $\gamma$, $\nabla p = c_S^2 \nabla \rho$, which
leads to the dispersion relation of MHD waves, expressed in terms of the
three variables $(\rho, {\bf v}, {\bf B})$, i.e., the mass density $\rho$,
the velocity ${\bf v}$, and the magnetic field strength ${\bf B}$. Linearizaton
of the equations for small perturbations in $({\rho}, {\bf v}, {\bf B})$
around a stationary solution, neglecting the gravity $g$, using the definition
of the Alfv\'en speed $v_A=B_0/\sqrt{4 \pi \rho_0}$, and inserting the
Fourier form ($d/dt\mapsto i\omega$, $\nabla \mapsto ik$) yields then the
well-known dispersion relation for magneto-acoustic waves,
\begin{equation}
	\omega^4 - k^2(c_S^2+v_A^2)\omega^2 + k_z^2k^2c_S^2v_A^2 = 0
	\ ,
\end{equation}
which has two branches of solutions, called fast and slow magneto-acoustic 
waves. For the non-magnetic case ($B=0$ and $v_A=0$), the phase speed is
equal to the sound speed, $v_{ph}=c_S$, which is the slow or acoustic mode,
producing a longitudinal wave, is non-dispersive, and compressional
($\rho_1 \propto v_1$). In a magnetized plasma ($B \neq 0$), the phase
speed depends on the angle between the magnetic field ${\bf B}$ and the 
wave propagation vector ${\bf k}$ and is called the fast mode. 
For the special case of a perpendicular angle $\theta=90^\circ$, called a
shear Alfv\'en wave, the phase speed is $v_{ph}=\sqrt{c_S^2 + v_A^2}$, 
and the perpendicular wave is dispersive and incompressible
($B_1 \propto v_1$). 

These basic properties tell us already how these waves are detected in
the solar corona. Acoustic waves show up as pressure fluctuations and
thus can be detected from density fluctuations. Optically thin emission
in EUV and soft X-rays has an emission measure that is proportional to
the square of the electron density, $EM \propto \int n_e^2(z) dz$, and
thus propagating sound waves can be traced from the associated emission
measure variations. A further proof of the identity of detected acoustic
waves is the measurement of the propagation speed, which turned out to be
identical to the sound speed $c_S \approx 150\ T_{MK}$ km s$^{-1}$ expected
for the observed temperature of the plasma (i.e., $T\approx 1.0$ MK at
171 \ang ). Such slow (acoustic) waves have been detected by 
\cite[DeForest \& Gurman (1998)]{D1}, 
\cite[Berghmans \& Clette (1999)]{B1}, 
\cite[DeMoortel et al.~(2000]{D2}; 
\cite[2002)]{D3}, and
\cite[Robbrecht et al.~(2001)]{R1}, 
using SOHO/EIT and TRACE (Table 1), and have even been re-discovered
in Yohkoh data
\cite[(Mariska 2005]{M9},
\cite[2006)]{M10}. 
These propagating acoustic waves were all detected in upward direction
in open magnetic field structures, where no reflection occurs.
In closed-field structures, acoustic waves seem
to be excited at one footpoint by some flare-like disturbance, which then
propagate with sound speed to the opposite footpoint and are reflected,
building up a standing (acoustic) oscillation or eigen-mode. Such acoustic 
oscillations, detected from the Dopplershift of the velocity disturbance 
as well as from the density variations, have been reported from SOHO/SUMER 
observations 
\cite[(Wang et al.~2002]{W1}; 
\cite[Kliem et al.~2002)]{K1}, 
with periods of order $P\approx 10-20$ min (Table 1).

\begin{table}
\begin{center}
\begin{tabular}{llll} 
\hline
MHD wave type     & Period range        & Observations  & References of examples \\ 
\hline
\underbar{\sl MHD Oscillations}&        &               &               \\
Fast kink mode    & $\approx 3-5$ min   & TRACE         & \cite[Aschwanden et al.~(1999)]{A1} \\
                  &                     &               & \cite[Nakariakov et al.~(1999)]{N1} \\
Fast sausage mode & $\approx 1-10$ s    & Radio         & \cite[Aschwanden (1987)]{A2} \\
                  &                     & Nobeyama      & \cite[Asai et al.~(2001)]{A3} \\
                  &                     & Nobeyama      & \cite[Melnikov et al.~(2002]{M1}, \cite[2005)]{M2} \\
Slow (acoustic) mode& $\approx 10-20$ min & SOHO/SUMER  & \cite[Wang et al.~(2002)]{W1} \\
                  &                     & SOHO/SUMER    & \cite[Kliem et al.~(2002)]{K1} \\
\hline
\underbar{\sl Propagating MHD Waves}&   Velocity range & &              \\
Slow (acoustic) waves & $75-150$ km/s   & SOHO/EIT      & \cite[DeForest \& Gurman (1998)]{D1} \\
                  & $75-200$ km/s       & SOHO/EIT      & \cite[Berghmans \& Clette (1999)]{B1} \\
                  & $70-235$ km/s       & TRACE         & \cite[DeMoortel et al.~(2000]{D1}, \cite[2002)]{D2} \\
                  & $65-150$ km/s       & TRACE, SOHO/EIT&\cite[Robbrecht et al.~(2001)]{R1} \\
Fast (Alfv\'enic) waves & $2100$ km/s   & SECIS         & \cite[Williams et al.~(2001)]{W2} \\
                  &                     &               & \cite[Katsiyannis et al.~(2003)]{K2} \\
		  & $1000-4000$ km/s 	& CMCP/NSO	& \cite[Tomczyk et al.~2007)]{T1}\\ 
Fast kink waves   & $100-500$ km/s      & TRACE         & \cite[Verwichte et al.~(2005)]{V1} \\
\hline
\end{tabular}
\caption{MHD wave types identified in the solar corona.}
\end{center}
\end{table}

The detection and identification of fast MHD waves is quite different from
acoustic modes because of their different physical properties. Let us
consider the fast MHD standing waves (or eigen-modes). The fast kink mode
is an asymmetric mode and shows perpendicular displacements of the loop
centroid, but only with relatively small amplitudes in the order of a few
percent of the loop length. Only in 1999, after the high-resolution EUV
images of TRACE with a spatial resolution of $\approx 1\arcsec$ became 
available, such fast kink modes were discovered 
\cite[(Aschwanden et al.~1999]{A1};
\cite[Nakariakov et al.~1999)]{N1}. 
Theoretically, these fast kink modes could have
been detected with SOHO/EIT earlier, but apparently the image cadence was
never chosen to be sufficiently fast to resolve the typical kink mode periods
of $P\approx 3-5$ min. 

The other branch of fast MHD modes is the sausage
mode, a symmetric mode that should show up as a radial oscillation, and thus
the emission measure should be modulated by the 4th power of the wave 
amplitude, i.e., $\Delta EM(t)/EM \propto 4 \Delta r(t)/r$. Although there
was a lot of evidence for fast MHD oscillations in terms of the expected
fast periods in the order of $P\approx 1-10$ s from non-imaging radio
observations (e.g., see review of 
\cite[Aschwanden 1987)]{A2}, there is still rather
sparse evidence from imaging observations. One case attributed to the
fast sausage mode has been reported from Nobeyama observations
\cite[(Asai et al.~2001]{A3}; 
\cite[Melnikov et al.~2002]{M1}, 
\cite[2005]{M2}; 
\cite[Nakariakov et al.~2003)]{N2}, 
based on periods from hard X-ray and microwave light curves and the 
estimated Alfv\'en crossing time. However, the emission measure modulation
and spatially-resolved radial oscillations of the thermal plasma associated
with the fast sausage MHD mode have never been directly observed. The
theory predicts an anti-correlated variation of the cross-sectional loop
radius with its electron density, which should be detectable with current
soft X-ray imagers (such as HINODE/XRT). Moreover, the fast sausage mode
is also subject to a wavenumber cutoff at a phase speed of $v_{ph}=v_{Ae}$
(when the phase speed matches the external Alfv\'en speed), which implies
relatively fat loops (which a large width/length ratio $w/l$) or very dense
loops. This requirement for the density contrast between the internal and 
the external density, $n_0/n_e$, 
\begin{equation}
	{n_0 \over n_e} \gapprox 2.334 \left({l \over w}\right)^2
	\ ,
\end{equation} 
is especially restrictive for slender loops (say $l/w > 10$), which
requires very overdense loops ($n_0/n_e > 230$) that only exist in
post-flare conditions 
\cite[(Aschwanden et al.~2004)]{A4}. The detection of such
fast sausage modes with periods of a few seconds from the thermal plasma
is still difficult because sub-second cadences of soft X-ray imagers are
rarely available. 

The detection of propagating fast Alfv\'enic waves is also difficult, because
a purely Alfv\'enic wave is incompressible and does not produce any
modulation of the density that could be easily observed. Alfv\'enic
waves cause a modulation $B_1 \propto v_1$, but the magnetic field
fluctuations $B_1$ cannot be measured by current methods, while 
fluctuations of velocities $v_1$ can only be measured from line broadening,
but are subject to interpretational ambiguities (such as in terms of
turbulence). However, there is a continuous range of intermediate
magneto-acoustic waves between the slow (acoustic) mode and the fast
(Alfv\'enic) mode, which display a mixed characteristic of compressible
and incompressible waves, which leaves a possibility for detection by
density modulations. Claims for detected fast Alfv\'enic waves in the 
solar corona come mainly from the phase speed of observed disturbances, 
which were found to be of order $v_{ph} \approx 2100$ km s$^{-1}$ from 
high-cadence optical images during a solar eclipse 
\cite[(Williams et al.~2001]{W2}; 
\cite[Katsiyannis et al.~2003)]{K2}, 
and of order $v_{ph} \approx 1000-4000$ km s$^{-1}$ 
from a {\sl Coronal Multi-Channel Polarimeter} at NSO 
\cite[(Tomczyk et al.~2007)]{T1}. 

There are also exciting observations of oscillations from flare stars.
\cite[Gary et al.~(1982)]{G1} reported oscillations with periods of $P=56$ s at
9.4 GHz from the flare star UV Ceti, which probably is coherent radio 
emission modulated by the magnetic field variations of a fast MHD mode.
There are reports of oscillations with periods of $P=9-246$ s in optical
wavelengths 
\cite[(Andrews 1990]{A5}; 
\cite[Mullan et al. 1992]{M3}; 
\cite[Mathioudakis et al.~2003)]{M5}
which are harder to explain because we do not understand how MHD modes
modulate optical emission. And there is a number of observed stellar
oscillations in soft X-rays with periods of $P=56-750$ s 
\cite[(Mullan \& Johnson 1995]{M4}; 
\cite[Mitra-Kraev et al.~2005)]{M6}, 
which have all faster periods
than what we are used to see for slow acoustic modes in the Sun, with
typical periods of $P\approx 10-20$ min, at a temperature of $T=7$ MK
measured with SUMER/SOHO. However, since stellar flares have significantly
higher temperatures, in the range of $T=10-100$ MK, the acoustic speed
is also faster, in the order of $c_S=450-1500$ km s$^{-1}$, which could
explain faster periods of standing acoustic waves.

Thus, our observations of MHD standing and propagating waves in the solar
corona become more complete, but some of the waves modes are still difficult
to detect due to insufficient cadence, spatial resolution, and sensitivity.
In stellar flares we lack the spatial resolution, and thus cannot determine
the loop length that is important to infer the acoustic or Alfv\'enic loop
crossing time, and thus have to rely on applications of scaling laws from
solar flares. For a review of other recent work on the topic of waves
in coronae (omitted here) see also review by Nakariakov in this book.

\section{Turbulence in Coronae}

Convection and turbulence are important in fluids with high Reynolds numbers. 
Since the {\sl magnetic Reynolds number} $R_m = l_0 v_0/{\eta}_m$ (or 
{\sl Lundquist number}) is high in the coronal plasma ($R_m \approx 
10^8-10^{12}$), turbulence may also develop in coronal loops (although
there is the question whether turbulence could be suppressed in coronal loops 
due to the photospheric line-tying). Theoretical models and numerical simulations
that study MHD turbulence include the {\sl kinematic viscosity} or 
{\sl shear viscosity} ${\nu}_{visc}$ in the MHD momentum equation, 
\begin{equation}
        \rho {D {\bf v} \over Dt}  = -\nabla p -\rho {\bf g} + ({\bf j} \times {\bf B})
        + {\nu}_{visc} \rho \left[ \nabla^2 {\bf v} + {1\over 3} \nabla (\nabla \cdot {\bf v}) \right]  \ ,
\end{equation}
and the {\sl magnetic diffusivity} ${\eta}_m = c^2/4 \pi \sigma$ in the 
MHD induction equation,
\begin{equation}
        {\partial {\bf B} \over \partial t}
        = \nabla \times ({\bf v} \times {\bf B})
        + {\eta}_m \nabla^2 {\bf B} \ .
\end{equation}
Similar to the models of stress-induced current cascades, random footpoint motion
is assumed to pump energy into a system at large scales (into eddies the size of 
a granulation cell, $\approx 1000$ km), which cascade due to turbulent motion 
into smaller and smaller scales, where the energy can be more efficiently 
dissipated by friction, which is quantified by the kinematic or shear viscosity
coefficient ${\nu}_{visc}$. Friction and shear are dynamical effects resulting 
from the nonlinear terms ($v_{1,i} v_{1,j}$), ($v_{1,i} B_{1,j}$), 
($B_{1,i} v_{1,j}$), and ($B_{1,i} B_{1,j})$ in Eqs.~(3.1-3.2) and are only 
weakly sensitive to the detailed dynamics of the boundary conditions. 
Analytical (3D) models of MHD turbulence have been developed by 
\cite[Heyvaerts \& Priest (1992)]{H1}, 
\cite[Inverarity et al.~(1995)]{I1},
\cite[Inverarity \& Priest (1995)]{I2}, and 
\cite[Milano et al.~(1997)]{M7}, 
where the nonlinear viscosity terms are specified as diffusion coefficients. 
These turbulent diffusion coefficients are free parameters, which are constrained 
self-consistently by (1) assuming that the random footpoint motion has a
turbulence power spectrum (e.g., a Kolmogorov spectrum $P(k) \propto k^{5/3}$); 
and (2) by matching the observed macroscopic parameters (i.e., velocity of 
footpoint motion, density, and magnetic field). 
\cite[Heyvearts \& Priest (1992)]{H1}
predict turbulent velocities of ${v}_{turb} \approx 20-30$ km s$^{-1}$, which 
are consistent with the excess broadening of lines observed with SUMER,
which shows a peak of $\xi = 30$ km s$^{-1}$ at a transition region 
temperature of $T \approx 3 \times 10^5$ K (e.g., 
\cite[Chae et al.~1988)]{C1}.

Analytical models of turbulent heating are applied to sheared arcades 
\cite[(Inverarity et al.~1995)]{I1}
and twisted fluxtubes 
\cite[(Inverarity \& Priest 1995)]{I2}. Turbulent heating
has been numerically simulated in a number of studies, which exhibit a high 
degree of spatial and temporal intermittency 
\cite[(Einaudi et al.~1996a]{E1},
\cite[b]{E2}; 
\cite[Dmitruk \& Gomez 1997)]{D4}.
Such simulations show how larger eddies fragment into smaller ones, 
forming current sheets and triggering magnetic reconnection during this process. 
Heating occurs by Ohmic dissipation in the thinnest current sheets. 
\cite[Milano et al.~(1999)]{M8} 
emphasize that the locations of heating events coincide with quasi-separatrix 
layers. The formation of such current sheets
has also been analytically studied in the context of turbulent heating by 
\cite[Aly \& Amari (1997)]{A6}.
Numerical simulations reveal intermittent heating
events with energies of $E_H=5\times 10^{24}$ to $10^{26}$ erg and a frequency 
distribution with a powerlaw slope of $\alpha \approx 1.5$, similar to observed 
nanoflare distributions in EUV 
\cite[(Dmitruk \& Gomez 1997]{D4}; 
\cite[Dmitruk al.~1998)]{D5}.

\bigskip
Although we have a lot of theoretical modeling and MHD simulations on the
fundamental process of turbulence, the major question we face is what is
the observational evidence that this process operates in the solar corona,
and how can it be tested observationally. One of the key observables that
is thought to provide evidence for turbulence is the nonthermal line
broadening in EUV and soft X-ray lines. However, the detection of nonthermal
line broadening is not a sufficient criterion, because it can also be caused
by unresolved flows that are not necessarily due to turbulence. An additional
constraint could be the altitude dependence of the nonthermal excess line
broadening $\xi(T)$, which was found to increase from $\xi \approx 5$ km s$^{-1}$
in the temperature minimum region ($T\approx 10^4$ K) to a maximum of
$\xi \approx 30$ km s$^{-1}$ in the transition region ($T \approx 3 \times
10^5$ K), and to decrease further upwards in the lower corona (e.g., 
\cite[Chae et al.~(1998)]{C1}. 
Equating the heating rate to the turbulent dissipation rate,
\cite[Chae et al.~(1998)]{C1} was able to reproduce the observed height
and temperature dependence of the nonthermal line broadening. Implicitly,
this model implies, however, that the heating rate is concentrated in the
transition region and becomes negligible in the coronal part of loops.
The hypothesis of footpoint heating, moreover, is also supported by at least
ten other observational constraints 
\cite[(Aschwanden 2001]{A7},
\cite[Aschwanden et al.~2007)]{A8}. 

On the other side, the identification of turbulence has been clearly established 
in the outer or extended corona. The solar wind is considered as a medium 
with well-developed MHD turbulence, as shown by the coupled variations of
the three magnetic field components and the fluid velocity
\cite[(Belcher \& Davis 1971)]{B2}, which is expected for Alfv\'en waves
($v_1 \propto B_1$). In the framework of the ``MHD turbulent cascade",
energy is injected at large scales and converted into smaller scales
(or eddies) with subsequent dissipation at the smallest scales, resulting
into plasma heating 
\cite[(Goldstein et al.~1995)]{G2}.
The inertial range of the turbulent power spectrum has a powerlaw slope
of 3/2 or 5/3 (Kraichnan or Kolmogorov spectrum), as it is observed in
the interplanetary magnetic field
\cite[(Leamon et al.~1998)]{L1}. 
The current thinking is that the energy input into the fast solar wind 
comes from kink-mode field motion, generated by transverse shaking in
intergranular lanes, which is then transformed into Alfv\'en waves in the 
canopies of the transition region and propagates into the fast solar wind.
A new insight is that the upward propagating Alfv\'en waves become partially
reflected in the solar wind, so that MHD turbulence develops from the
nonlinear interaction of outward and inward propagating Alfv\'en waves
\cite[(Cranmer \& van Ballegooijen 2005)]{C2}.
The heating of the solar wind is then accomplished by ion-cyclotron
wave-particle interactions of these Alfv\'en waves, as it has been
convincingly demonstrated by the surprisingly large ion temperatures,
ion outflow speeds (H$^\circ$, O$^{5+}$), and velocity distribution
anisotropies of positive ions observed by SOHO/UVCS above coronal holes
in the solar wind at distances of $\approx 1-3$ solar radii
\cite[(Li et al.~1998)]{L2}.

In summary, in the solar context, turbulence seems to play an important
role in the transition region and in the extended corona (solar wind), 
while there is less evidence for heating by turbulence in the closed-field 
corona.

\begin{table}
\begin{center}
\begin{tabular}{lll}
\hline
{\sl Physical aspect:}          & {\sl Observational signature:}        & {\sl References:}     \\
\hline
X-point geometry                & Cusp in LDE events                    & \cite[Tsuneta et al.~(1992)]{T2}\\
\hline
X-point altitude                & Time-of-flight measurements            & \cite[Aschwanden et al.~(1996)]{A13}\\
                                & Above-the-looptop HXR sources         & \cite[Masuda et al.~(1994)]{M11}\\
\hline
X-point rises with time         & Increasing footpoint separation       & \cite[Sakao et al.~(1998)]{S4}\\
                                & or double-ribbon separation           & \cite[Fletcher \& Hudson (2001)]{F2}\\
\hline
X-point symmetry, horizontal    & Simultaneous HXR emission             & \cite[Sakao et al.~(1994)]{S5}\\
                                & at conjugate footpoints               &                       \\
\hline
X-point symmetry, vertical      & Bi-directional type III bursts        & \cite[Aschwanden et al.~(1995)]{A14}\\
                                & Coincidence HXR + type III            & \cite[Aschwanden et al.~(1993)]{A15}\\
				& Dual coronal HXR sources		& \cite[Sui \& Holman (2003)]{S3}\\
\hline
Post-reconnection relaxation    & Loop shrinkage ratio                  & \cite[Forbes \& Acton (1996)]{F3}\\
                                & cooling loops below hot loops         & \cite[Svestka et al.~(1987)]{S6}\\
\hline
Quadrupolar geometry            & Interacting flare loops                & \cite[Hanaoka (1996]{H3},\cite[1997)]{H4}\\
                                &                                       & \cite[Nishio et al.~(1997)]{N3}\\
                                &                                       & \cite[Aschwanden et al.~(1999a)]{A16}\\
\hline
3D nullpoint geometry           & Fan dome and spine morphology         & \cite[Fletcher et al.~(2001)]{F4}\\
\hline
Reconnection inflows            & EUV inward motion                      & \cite[Yokoyama et al.~(2001)]{Y1}\\
\hline
Reconnection outflows           & Supra-arcade downflows                  & \cite[McKenzie \& Hudson (1999)]{M12}\\
\hline
Slow-mode standing shocks       & High-temperature ridges               & \cite[Tsuneta (1996)]{T3}\\
\hline
Fast-mode standing shocks       & High density above looptop            & \cite[Tsuneta et al.~(1997)]{T4}\\
                                & Above-the-looptop HXR                 & \cite[Masuda et al.~(1994)]{M11}\\
\hline
Plasmoid ejection               & Upward-moving plasmoid                & \cite[Shibata et al.~(1992)]{S7}\\
                                & Streamer blobs                        & \cite[Sheeley et al.~(1997)]{S8}\\
\hline
Downward conduction             & Downward thermal fronts               & \cite[Rust et al.~(1985)]{R2}\\
\hline
Chromospheric evaporation       & Line broadening                       & \cite[Antonucci et al.~(1986)]{A17}\\
                                & SXR upflows                            & \cite[Acton et al.~(1982)]{A18}\\
                                & SXR blueshifts                        & \cite[Czaykowska et al.~(1999)]{C3}\\
                                & H$\alpha$ redshifts                   & \cite[Zarro \& Canfield (1989)]{Z1}\\
                                & Momentum balance                      & \cite[Wuelser et al.~(1994)]{W3}\\
                                & HXR/H$\alpha$ ribbons                 & \cite[Hoyng et al.~(1981)]{H5}\\
\hline
\end{tabular}
\caption{Key observations that provide evidence for magnetic reconnection in
solar flares and CMEs 
\cite[(Aschwanden 2004)]{A0}.}
\end{center}
\end{table}

\section{Magnetic Reconnection in Coronae}

Theory and numerical simulations of magnetic reconnection processes
in the solar corona have been developed for steady 2D reconnection,
bursty 2D reconnection, and 3D reconnection. Only steady 2D
reconnection models can be formulated analytically, which provide basic
relations for inflow speed, outflow speed, and reconnection rate, but represent
oversimplifications for most (if not all) observed flares. A more realistic
approach seems to be bursty 2D reconnection models, which involve the
tearing-mode and coalescence instability and can reproduce the sufficiently fast
temporal and small spatial scales required by solar flare observations.
The sheared magnetic field configurations and the existence or coronal and
chromospheric nullpoints, which are now inferred more commonly in
solar flares, require ultimately 3D reconnection models,
possibly involving nullpoint coalescence, spine reconnection, fan reconnection,
and separator reconnection. Magnetic reconnection operates in two
quite distinct physical parameter domains (in collisional or collisionless plasma): 
(i) in the chromosphere during
magnetic flux emergence, magnetic flux cancellation, and so-called explosive
events, and (ii) under coronal conditions during microflares, flares,
and CMEs. The best known flare/CME models entail magnetic reconnection
processes that are driven by a rising filament/prominence, by flux emergence, by
converging flows, or by shear motion along the neutral line. Flare
scenarios with a driver perpendicular to the neutral line (rising prominence,
flux emergence, convergence flows) are formulated as 2D reconnection models
\cite[(Kopp \& Pneuman 1976]{K3}; 
\cite[Heyvaerts et al.~1977]{H2}; 
\cite[Forbes \& Priest 1995]{F1}; 
\cite[Uchida 1980)]{U1},
while scenarios that involve shear along the neutral line
(tearing-mode instability, quadrupolar flux transfer, the magnetic breakout model,
sheared arcade interactions) require 3D descriptions 
\cite[(Sturrock 1966]{S1};
\cite[Antiochos et al.~1999]{A9}; 
\cite[Somov et al.~1998)]{S2}. 
Ultimately, most of these partial
flare models could be unified in a 3D model that includes all driver mechanisms.
Observational evidence for magnetic reconnection in flares includes the
3D geometry, reconnection inflows, outflows, detection of shocks, jets, ejected
plasmoids, and secondary effects like particle acceleration, conduction
fronts, and chromospheric evaporation processes (summarized in Table 2). 
Magnetic reconnection not only operates locally in flares, it also organizes 
the global corona by large-scale restructuring processes.

Let us mention a few recent studies that stimulate and challenge our
thinking about magnetic reconnection processes in solar flares. Dual coronal
hard X-ray sources with vertically symmetric energy sources have been discovered
with RHESSI, which for the first time directly demonstrate the vertical symmetry 
of an X-type reconnection geometry 
\cite[(Sui \& Holman 2003)]{S3}. 
A challenging phenomenon that has been observed in a few flares is the initial
altitude drop of the coronal hard X-ray source before it rises in the later
flare phase
\cite[(Sui \& Holman 2003]{S3}; 
\cite[Asai et al.~2004)]{A10}. 
Is it related to the relaxation of newly-reconnected field lines
or to the local implosion after a CME launch? The spatial evolution of reconnection
in a flare arcade was found to be quite complex: sometimes it can be tracked from
footpoint motions that increase in separation and move systematically along the ribbons 
\cite[(Krucker et al.~2003)]{K4}, and
sometimes they move along ribbons rather than apart as predicted by the Kopp-Pneuman model
\cite[(Grigis \& Benz 2005)]{G3}. 
Important insight into the reconnection process was also found from determining
the reconnection rate via the footpoint velocity and the local footpoint magnetic field 
\cite[(Asai et al.~2004a)]{A12}.
The energy release rate can then be obtained from the product of the Poynting flux 
and the area of the reconnection region ($A$), 
\begin{equation}
	{dE \over dt} = 2 {B_c^2 \over 4\pi} v_i A \ ,
\end{equation}
which was found to correlate with the hard X-ray flux
\cite[(Asai et al.~2004a]{A12};
\cite[Krucker et al.~2005)]{K5}.
RHESSI continues to provide important physical parameters to disentangle the
spatial and temporal evolution of magnetic reconnection processes.

\section{Particle Acceleration in Coronae}

Particle acceleration in solar flares is mostly explored by theoretical models,
because neither macroscopic nor microscopic electric fields are directly measurable
by remote-sensing methods. The motion of particles can be described in terms of
acceleration by parallel electric fields, drift velocities caused by perpendicular
forces (i.e., $E\times B$-drifts), and gyromotion caused by the Lorentz force of the
magnetic field. Theoretical models of particle acceleration in solar
flares can be broken down into three groups: (1) DC electric field acceleration,
(2) stochastic or second-order Fermi acceleration, and (3) shock acceleration
(for an overview see Table 3; for references see 
\cite[Aschwanden 2004, p.470)]{A0}.
In the models of the first group, there is a paradigm shift from
large-scale DC electric fields (of the size of flare loops) to small-scale
electric fields (of the size of magnetic islands produced by the tearing mode
instability). The acceleration and trajectories of particles is studied more
realistically in the inhomogeneous and time-varying electromagnetic fields
around magnetic X-points and O-points of magnetic reconnection sites, rather
than in static, homogeneous, large-scale Parker-type current sheets.
The second group of models entails stochastic acceleration by gyroresonant
wave-particle interactions, which can be driven by a variety of electrostatic
and electromagnetic waves, supposed that wave turbulence is present at a
sufficiently enhanced level and that the MHD turbulence cascading process is
at work. The third group of acceleration models includes a rich variety
of shock acceleration models, which is extensively explored in magnetospheric
physics and could cross-fertilize solar flare models. Two major groups of models
are studied in the context of solar flares (i.e., first-order Fermi acceleration
or shock-drift acceleration, and diffusive shock acceleration).

\begin{table}
\begin{center}
\begin{tabular}{ll}
\hline
Acceleration Mechanisms                         &Electromagnetic fields \\
\hline
{\sl {DC electric field acceleration}}:                 	 &       \\
    $-$ Sub-Dreicer fields, runaway acceleration 	         &$E < E_D$\\
    $-$ Super-Dreicer fields 	                                 &$E > E_D$\\
    $-$ Current sheet (X-point) collapse                        &$E = -u_{inflow}\times B$\\
    $-$ Magnetic island (O-point) coalescence                   &$E_{conv} = -u_{coal} \times B$\\
    $-$ (Filamentary current sheet: X- and O-points)&  \\
    $-$ Double layers                                           &$E =-\nabla V$\\
    $-$ Betatron acceleration (magnetic pumping)                &$\nabla \times E = -(1/c)(dB/dt)$\\
\hline
{\sl {Stochastic (or second-order Fermi) acceleration}}:        &       \\
    Gyroresonant wave-particle interactions (weak turbulence) with: &   \\
    $-$ whistler (R-) and L-waves                               &$k \parallel B$\\
    $-$ O- and X-waves                                          &$k \perp B$\\
    $-$ Alfv\'en waves (transit time damping)                   &$k \parallel B$\\
    $-$ Magneto-acoustic waves                                  &$k \perp B$\\
    $-$ Langmuir waves                                          &$k \parallel B$\\
    $-$ Lower hybrid waves                                      &$k \perp B$\\
\hline
{\sl {Shock acceleration}}:                                	&   \\
    Shock-drift (or first-order Fermi) acceleration              &   \\
    $-$ Fast shocks in reconnection outflow                      &   \\
    $-$ Mirror-trap in reconnection outflow                      &   \\
    Diffusive-shock acceleration                                &   \\
\hline
\end{tabular}
\caption{Overview of particle acceleration mechanisms in solar flares (Aschwanden 2004).}
\end{center}
\end{table}

New aspects are that shock acceleration is now applied to the outflow regions of
coronal magnetic reconnection sites, where first-order Fermi acceleration at the
standing fast shock is a leading candidate. Traditionally, evidence for
shock acceleration in solar flares came mainly from radio type II bursts. New
trends in this area are the distinction of different acceleration sites that
produce type II emission: flare blast waves, the leading edge of CMEs (bowshock),
and shocks in internal and lateral parts of CMEs. In summary we can say
that (1) all three basic acceleration mechanisms seem to play a role to a variable
degree in some parts of solar flares and CMEs, (2) the distinction between
the three basic models become more blurred in more realistic (stochastic) models, and
(3) the relative importance and efficiency of various acceleration models can
only be assessed by including a realistic description of the electromagnetic
fields, kinetic particle distributions, and MHD evolution of magnetic
reconnection regions pertinent to solar flares.

Particle kinematics, the quantitative analysis of particle trajectories, has been
systematically explored in solar flares by performing high-precision energy-dependent
time delay measurements with the large-area detectors of the {\sl Compton Gamma-Ray
Observatory (CGRO)}. There are essentially five different kinematic processes that play a
role in the timing of nonthermal particles energized during flares: (1) acceleration,
(2) injection, (3) free-streaming propagation, (4) magnetic trapping, and (5)
precipitation and energy loss. The time structures of hard X-ray and radio emission
from nonthermal particles indicate that the observed energy-dependent timing is
dominated either by free-streaming propagation (obeying the expected electron
time-of-flight dispersion) or by magnetic trapping in the weak-diffusion limit
(where the trapping times are controlled by collisional pitch angle scattering).
The measurements of the velocity dispersion from energy-dependent hard X-ray
delays allows then to localize the acceleration region, which was invariably
found in the cusp of postflare loops.

RHESSI observations produce new findings that challenge our previously established concepts.
For instance, 
the acceleration and/or propagation of flare-accelerated electrons and ions seems to have
distinctly different characteristics, since the 2.223 MeV neutron-capture gamma-ray line
was found to be significantly displaced from 200-200 keV hard X electrons 
\cite[(Hurford et al.~2003]{H6},
\cite[2006)]{H7}. Theoretical interpretations range from different acceleration path lengths
for electrons and ions in the stochastic acceleration process
\cite[(Emslie et al.~2004)]{E3}
to charge separation in the super-Dreicer electric field in a reconnecting non-neutral
current sheet 
\cite[(Zharkova \& Gordovskyy 2004)]{Z2}.

\section{Conclusions}

Our insights into the fundamental physical processes in the solar corona is exponentially
growing thanks to all the new high-resolution imaging and spectral observations. We have
identified almost all principal MHD wave modes in the solar corona, and theoretical studies
are exploring now second-order effects, which provide previously unknown physical parameters. 
Turbulence is a process that is established in the lower solar atmosphere (driven by the
subsurface magneto-convection), as well as in the solar wind (caused by interacting
outgoing and reflected Alfv\'en waves), but seems to be less important in the closed-field
corona (due to the large dissipation length of Alfv\'en waves). Magnetic reconnection is
an ubiquitous process in the solar corona, ranging from small-scale phenomena in the
transition region (explosive events, nanoflares, etc) to catastrophic large-scale
reconfigurations during flares and CMEs, but modeling problems deal mostly with the
relative importance of the drivers (magnetic flux emergence, cancellation, loss-of-equilibrium,
magnetic break-out, etc). The process of particle acceleration is, due to its microscopic physics,
the most difficult to study, and most conclusions are drawn from secondary processes, such as
the collisional thin- and thick-target interactions of the accelerated particles, or in-situ
particle detections in interplanetary space. It is imperative to get a grip on the
3D geometry and topology of magnetic reconnection regions and shock waves in order to get 
a glimpse on the electromagnetic fields that accelerate high-energy particles. It is clear
that only multi-wavelength observations with comprehensive modeling can lead to a deeper
physical understanding and to more reliable conclusions about these fundamental physical
processes in the solar corona. Conclusions about the operation of the same physical processes
in stellar coronae can only be drawn by inference, but the sensitivity threshold of stellar 
observations generally imply a bias towards more energetic flare processes.

\begin{acknowledgments}
Part of this work was supported by NASA contracts 
NAG5-13490 (Living with a Star Targeted Research \& Technology),
NNG06GC25G (Solar and Heliospheric Physics),
NAS5-38099 (TRACE mission),
and NAS5-98033 (RHESSI mission, through University of California, Berkeley,
subcontract SA2241-26308PG). 
\end{acknowledgments}

\end{document}